\documentclass[a4paper,fleqn]{cas-sc}
\usepackage[authoryear,longnamesfirst]{natbib}

\def\tsc#1{\csdef{#1}{\textsc{\lowercase{#1}}\xspace}}
\tsc{WGM}
\tsc{QE}
\tsc{EP}
\tsc{PMS}
\tsc{BEC}
\tsc{DE}

\newcommand{\ignore}[1]{}
\usepackage{array,multirow}
\usepackage{graphicx}
\usepackage{subfig}
\usepackage{graphicx}
\usepackage{multirow}
\usepackage[normalem]{ulem}
\useunder{\uline}{\ul}{}
\usepackage{bbding}
\usepackage{appendix}
\usepackage{color}
\usepackage{amsmath}

\begin{document}
\let\WriteBookmarks\relax
\def\floatpagepagefraction{1}
\def\textpagefraction{.001}

\shorttitle{Invariant Debiasing Learning for Recommender Systems}

\shortauthors{Bai et~al.}

\title [mode = title]{Invariant Debiasing Learning for Recommendation via Biased Imputation}                      
\tnotemark[1]

\tnotetext[1]{This work was supported by the National Key Research and Development Program of China (No.2023YFC3303800).}

\author{Ting Bai}






\affiliation{organization={Beijing University of Posts and Telecommunications},
    addressline={Xitu Cheng Road}, 
    city={Beijing},
    postcode={100876}, 
    country={China}}

\author{Weijie Chen}
\author{Cheng Yang}

\author{Chuan Shi*}
\cortext[cor1]{Corresponding author}


\begin{abstract}
Previous debiasing studies utilize unbiased data to make supervision of model training. They suffer from the high trial risks and experimental costs to obtain unbiased data. Recent research attempts to use invariant learning to detach the invariant preference of users for unbiased recommendations in an unsupervised way. However, it faces the drawbacks of low model accuracy and unstable prediction performance due to the losing cooperation with variant preference. In this paper, we experimentally demonstrate that invariant learning causes information loss by directly discarding the variant information, which reduces the generalization ability and results in the degradation of model performance in unbiased recommendations. Based on this consideration, we propose a novel lightweight knowledge distillation framework (\textbf{KD-Debias}) to automatically learn the unbiased preference of users from both invariant and variant information.
Specifically, the variant information is imputed to the invariant user preference in the distance-aware knowledge distillation process. Extensive experiments on three public datasets, i.e., Yahoo!R3, Coat, and MIND, show that with the biased imputation from the variant preference of users, our proposed method achieves significant improvements with less than $50\%$ learning parameters compared to the SOTA unsupervised debiasing model in recommender systems.  Our code is publicly available at https://github.com/BAI-LAB/KD-Debias.
\end{abstract}


\maketitle

\section{Introduction}
As recommender systems have achieved great success in providing personalized service, numerous studies focus on improving the accuracy of model performance~\citep{luo2024unbiased,wei2024llmrec,song2024xgcn,hua2024up5}. Recently, increasing attention has been paid to the long-term impacts on user experience and retention in commercial platforms, especially in reducing the risks brought by multiple kinds of bias, including the exposure bias~\citep{yang2018unbiased}, position bias~\citep{collins2018study}, popularity bias~\citep{chen2020esam,krishnan2018adversarial} and so on.
For example, popularity bias refers to popular items being overly recommended, which is at the expense of damages in recommendation ecology due to the under-recommendation of less popular items that users may be interested in. 
\begin{figure*}[htbp]
\centering
\subfloat[Performance Stability]{\includegraphics[width=0.24\textwidth]{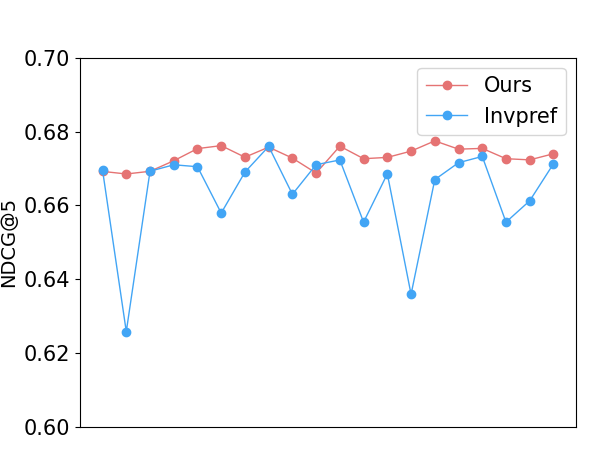}}
\subfloat[Distribution of preferences]{\includegraphics[width=0.24\textwidth]{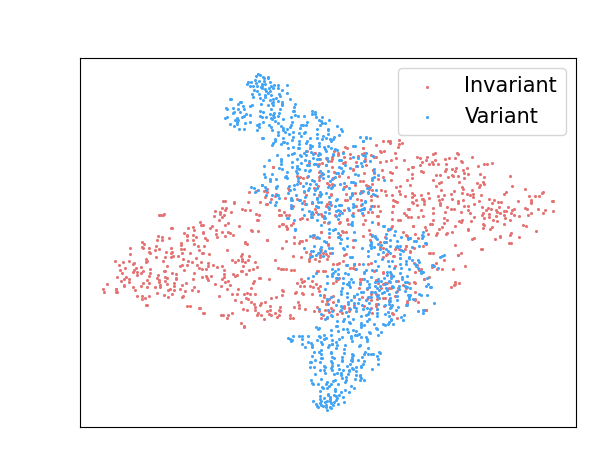}}
\subfloat[Information Entropy]{\includegraphics[width=0.24\textwidth]{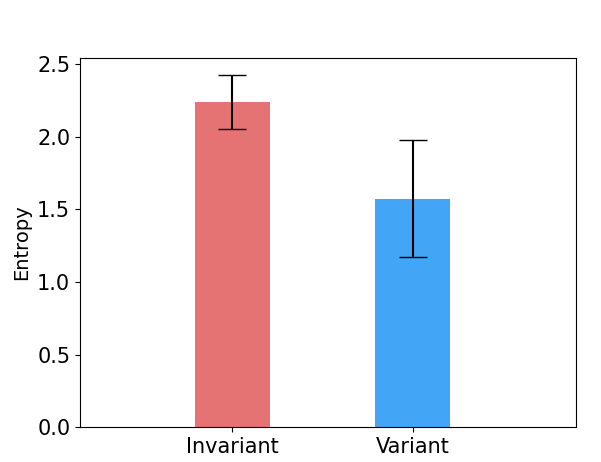}}
\subfloat[Accuracy Comparion]{\includegraphics[width=0.24\textwidth]{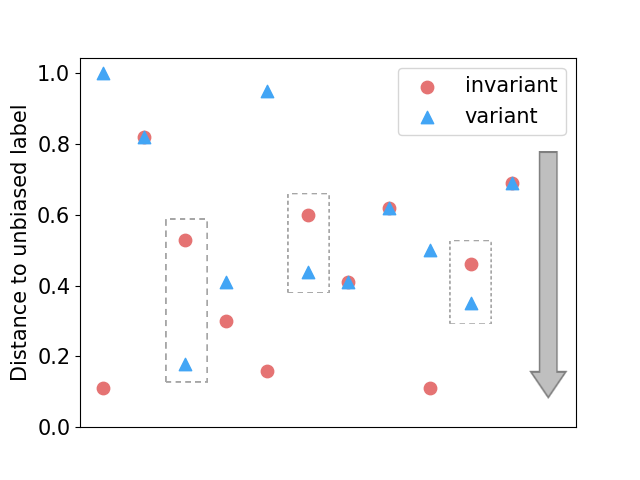}}
\caption{ The drawbacks (i.e.,  low model accuracy and unstable prediction ability) of the invariant learning method InvPref which solely relying on invariant information to make unbiased recommendations.
(a) The model performances of our model KD-Debias and InvPref on the Yahoo!R3 dataset with random seeds in the 20 repeated experiments,the horizontal axis represents different experimental groups; (b) The distributions of invariant and variant preference in InvPref; (c) The information entropies of invariant and variant preference; (d) The distance between the unbiased true label and the predicted label from invariant and variant preference: the smaller the distance is, the more accurate the prediction is. The cases in which the prediction from variant preference is more accurate to the unbiased true label are circled with the dotted line, implying their potential utility to enhance the model performance, the horizontal axis refers to different data samples. }
\label{fg:findings}
\end{figure*}
The studies of debiasing in recommender systems can be generally classified into two basic routes: one is to address a specific bias, like popularity bias~\citep{chen2020esam,krishnan2018adversarial} or
position bias~\citep{collins2018study},
the other is general debiasing with various biases~\citep{li2024removing,li2023causal,zhang2024general}. 
In practice, in real recommendation scenarios, various types
of biases are often mixed up, for example, a user’s click behavior may be influenced by both popularity bias and exposure bias.

To deal with the mixed-up biases, general debiasing methods~\citep{li2024removing,li2023causal,zhang2024general} are proposed, and they usually require the supervision of unbiased data in the training process. However, the acquisition of such unbiased data in real recommender systems often entails considerable trial risks and human costs.
In response, recent studies attempt to apply invariant learning~\citep{creager2021environment,du2022invariant} to detach the invariant preference of users for unbiased recommendations in an unsupervised way (i.e., training without unbiased data). 
The debiasing method~\citep{wang2022invariant,tang2024unbiased} via invariant learning assumes that 
the observational user behaviors are determined by invariant preference (i.e. a user’s true preference) and the variant preference (affected by some unobserved confounders)\footnote{ In the debiasing model~\citep{wang2022invariant}: the invariant preference refers to the unbiased preference, and the variant preference refers to the biased preference affected by unobserved confounders.}.
They believe that the unobserved confounders may cause various biases, and only the invariant preference should be used for unbiased recommendations.
Conducted in an unsupervised way, such a debiasing model~\citep{wang2022invariant} avoids the high risks in collecting unbiased data.

However, we find it faces the drawbacks of low model accuracy and unstable prediction ability.
Fig.1 (a) illustrates the utilization of solely the invariant preference for generating unbiased recommendations across various random initialized seeds. 
Notably, the performance of the invariant learning model~\citep{wang2022invariant} (marked in blue line) exhibits marked instability, indicative of its limited generalization capability.
To figure out the reasons, we characterize the distributions of invariant preference and the variant preference via t-SNE algorithm~\citep{van2008visualizing}, as well as their information entropy in Fig.1 (b-c). We find out that the distribution of the invariant and variant preference are distinct different (subfigure b), and both of them contain abundant information (subfigure c). 
Besides, to verify whether the variant preferences are useless for debiasing learning, we evaluate their capability to make unbiased recommendations, as shown in Fig.1 (d): for a certain number of cases ( circled with the dotted line), the prediction from variant preference is more accurate to the true unbiased label.
This indicates that solely relying on invariant information to make unbiased recommendations will result in information loss, which degrades model performance and reduces its generalization ability. 
The presence of abundant information within users' variant preferences implies their potential utility in enhancing the effectiveness of unbiased recommendations. 
Based on the above consideration, our aim is to take full advantage of the variant information to enhance unbiased recommendations.

To tackle this challenge, we propose a novel distance-aware knowledge distillation framework (\textbf{KD-Debias}) to adaptively fuse the invariant and variant information for general debiasing. 
The knowledge distillation process provides a way to distill useful information from the variant information, which is automatically imputed to the unbiased data distribution
following the rules: 
comparing to the distribution of invariant information, the more different the distribution of the variant information is (i.e., greater distance), the more attention will be paid to the variant preference in the knowledge distillation process. 
Then a simple lightweight matrix factorization (MF) based student model is designed to capture the collaborative signals. 
With the soft predicted label from the teacher model, the MF based student model not only improves the model efficiency by reducing the number of parameters, but also enhance its generalization ability for the final unbiased predictions.
Our contributions are as follows:
\begin{itemize}
    \item We use invariant learning to conduct general debiasing with biased observational data only and demonstrate the necessity to incorporate the biased imputation from variant information in unbiased recommendations.
    \item We propose a lightweight distance-aware knowledge distillation framework, termed as KD-Debias, to automatically fuse the invariant and variant information, the distillated student model achieves the best model performance with less than $50\%$ learning parameters compared to the SOTA unsupervised debiasing method (i.e., InvPref).
    \item  We conduct extensive experiments on three public datasets, i.e., Yahoo!R3, Coat, and Mind, to show that with the imputation from biased information, our method achieves the best model performance in unbiased recommendations.
\end{itemize}

\section{Related Work}
In this section, we introduce the concepts related to our study briefly, including debiasing learning, invariant learning, and knowledge distillation methods in recommender systems.
\subsection{Debiasing in recommender systems} 
There are various kinds of biases in recommender systems, 
such as selection bias~\citep{li2023causal,saito2020asymmetric,li2022stabilized}, exposure bias~\citep{mansoury2022understanding,ning2023input}, popularity bias~\citep{zhao2022popularity,chen2023bias}, and other undefined biases caused by latent confounders~\citep{nam2020learning,wang2022unbiased,zhu2024mitigating}. 
The biases would hurt the experience of users and the revenues of platforms. 
As increasing attention has been paid to debiasing in recommender systems, more and more debiasing methods have been proposed~\citep{zhao2023mdi,wei2021model,qiu2021causalrec}.  
The studies of debiasing in recommender systems can be generally classified into two basic routes: one is to address a specific bias, like the exposure bias~\citep{du2023sequential,yang2018unbiased}, position bias~\citep{collins2018study,wu2024exploring}, popularity bias~\citep{chen2020esam,krishnan2018adversarial} and so on. 
For example, KDRec~\citep{yang2023debiased} introduces contrastive learning to capture item transition patterns and address popularity bias. Some researchers design inverse propensity score estimators to eliminate biases across different domains~\citep{li2021debiasing}.
The other is general debiasing with various biases mixed up~\citep{guo2024disentangled,zhang2024general,yang2024disentangled}.
For example, the IPS-based debiasing methods~\citep{schnabel2016recommendations,zhu2020unbiased} address biases by re-weighting the samples to make the biased distribution of training data approximate the distribution of unbiased data, 
iDCF~\citep{zhang2023debiasing} applies proximal causal inference
to infer the unmeasured confounders and identify the counterfactual feedback.
DIB~\citep{liu2023debiased} alleviates the confounding bias via information bottleneck.
AdvDrop~\citep{zhang2024general} employs adversarial learning to split the neighborhood into bias-mitigated interactions and bias-aware interactions for graph-based recommendations.
Despite their effectiveness, most of these methods train models under the supervision of unbiased data which is usually generated with high experimental costs, besides, the imbalance in the quantity of unbiased and normal data also raises concerns about overfitting issues.

Recently, some researchers have solved the debiasing issue in unsupervised ways~\citep{wang2020information,wang2022invariant}, i.e., training without unbiased data. 
For example, CVIB~\citep{wang2020information}  builds a counterfactual variational information bottleneck, mitigating bias by balancing learning between the factual and counterfactual domains; InvPref~\citep{wang2022invariant} captures the biases through estimating pseudo environments
labels and generate unbiased user preferences with invariant learning. Despite their ability to eliminate the dependence on unbiased data, these methods still face the drawbacks of low and unstable model performance according to our findings in Section 1. 

\subsection{Invariant Learning}
Invariant learning~\citep{arjovsky2019invariant,creager2021environment} assumes that the observed data originate from multiple different environments, which caused the heterogeneity of the distribution of observed data.
Relying on the principle of invariance inherent in causality, invariant learning captures representations with invariant predictive ability across environments.
By capitalizing on the invariant relationship that exists between features and labels across diverse distributions, invariant learning can ensure generalization and disregard pseudo-correlations even when there are deviations in distribution. 
Drawing from these strengths, the concept of invariant learning has been extensively applied in recommendation tasks, extending its utility into the domain of debiasing recommendations as well. 
For example, InvPref~\citep{wang2022invariant} attributes the heterogeneity of environments will cause biases. They assume that the observational user behaviors are determined by invariant preference (i.e. a user’s true preference) and
the variant preference (affected by some unobserved environment confounders). They capture the invariant user preferences across environments for unbiased recommendations. 
Compared with the existing debiasing methods with invariant learning,
we not only focus on the generalizability of invariant preference but also the information loss exhibited by invariant learning when it comes to personalized recommendations.

\subsection{Knowledge Distillation}
The knowledge distillation framework~\citep{hinton2015distilling} can distill the knowledge from teacher models (an ensemble of cumbersome models) to the student model (a small model) by using the “soft targets or label” produced by the teacher model for training the student model. In addition to the information of positive samples, aligning with “soft targets” could also carry a lot of information from negative ones, which improves the generalization ability of the student model. Benefiting from these strengths, many researchers employ knowledge distillation into recommendation systems to lightweight the model~\citep{gou2021knowledge,liu2020general} and improve the generalization capabilities~\citep{kang2020rrd,tang2018ranking,zhu2022debias} by ensemble models at a fine-grained level.
Recently, some studies~\citep{ding2022interpolative,liu2022kdcrec, chen2023unbiased} attempt to use knowledge distillation to improve the debiasing ability in recommender systems. KDCRec~\citep{liu2022kdcrec} proposes a flexible knowledge distillation method with four types of distillation strategies from the teacher model to the student model.
InterD~\citep{ding2022interpolative} achieves competitive performance in both the norm test and debiasing test by distilling knowledge from a biased model and an unbiased model. 
However, both of them are trained under the supervision of unbiased data, which brings the high experimental costs and concerns of the overfitting problem. Different from the existing knowledge distillation approaches on debiasing learning, we aim to achieve unbiased recommendations without the participation of unbiased data.

\begin{figure*}
\centering
\includegraphics[width=0.9\linewidth]{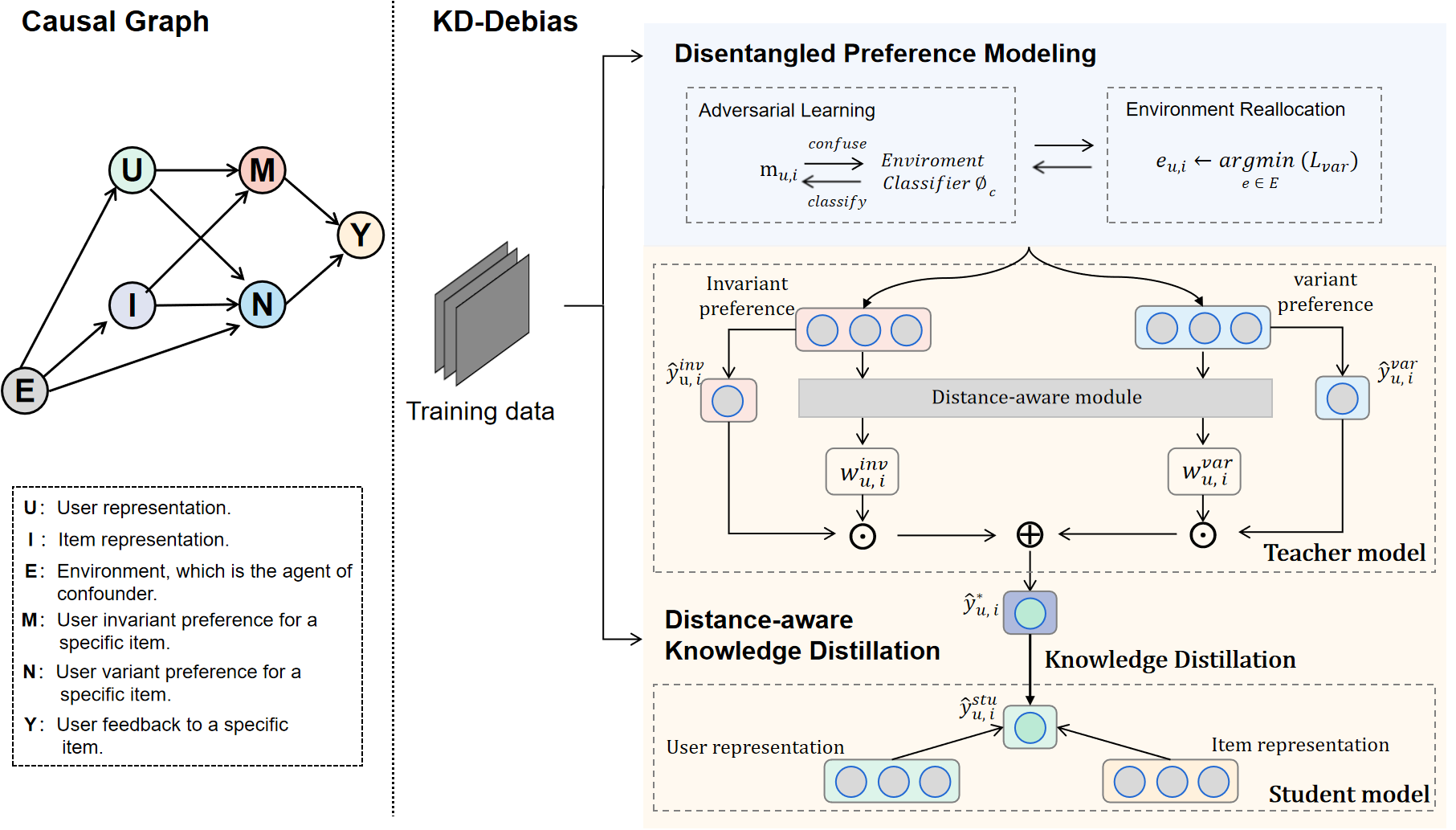}
  \caption{An Overview of KD-Debias framework.  It contains two components: the Disentangled Preference Modeling component and the Distance-aware Knowledge Distillation component. The disentangled preference modeling is conducted
on the left causal graph with an environment classifier to identify the biases from different environments. Then the knowledge distillation are conducted on the disentangled invariant preference and variant preference with a distance-aware fusion strategy.}
  \label{fig:model} 
\end{figure*}
\section{Our Proposed Method}
In order to make unbiased recommendations with the biased observational data alone, we use invariant learning to obtain the invariant and variant preferences of users in an unsupervised way. To fully make utilization of variant information, 
we propose a distance-aware knowledge distillation debiasing framework (KD-Debias) to adaptively fuse the invariant and variant information for debiasing in recommender systems.
KD-Debias consists of two components: i.e., the disentangled preference modeling component and the distance-aware knowledge distillation component. The illustration of our proposed framework is shown in Fig.~\ref{fig:model}.

\subsection{Disentangled Preference Modeling}
To learn the unbiased preference of users, we disentangle user preference into user invariant preference and variant preference. Following the method in invariant learning~\citep{wang2022invariant}, which assumes that the variant information is influenced by different latent confounders from environments and is regarded as biased information. 
Specifically, the disentangled preference modeling is conducted on a causal graph (see in the left part of Fig.~\ref{fig:model}) that describes the generation of observational user behaviors. Let $U$ and $I$ represent the set of users and items respectively. The preference of users $U$ towards items $I$ can be disentangled into invariant preference $M$ and variant preference $N$. Different from the invariant preference $M$, the variant preference $N$ is also directly influenced by the latent confounders in the environment $E$. The observed user feedback $Y$ is inevitably biased since it is affected by both invariant and variant preference. 

\subsubsection{Basic Summarization.}
To disentangle the invariant and variant preferences, 
invariant debiasing learning is conducted by an adversarial network. It designs an environment classifier to identify the biases from different environments. 
The invariant preference tries to confuse the environment classifier, while the classifier aims to identify the environment label of the invariant preference. 
A well-trained classifier can easily predict the environment label of the biases mixed in invariant preference. A well-optimized invariant user preference contains no bias that increases the difficulty of the classifier in identifying which environment the bias comes from.
Through the joint optimizations of the invariant preference and environment classifier, we can obtain the disentangled invariant and variant preferences of users.
Specifically, invariant learning can be divided into two steps, i.e., the invariant and variant preference modeling and the adversarial learning with the environment classifier.

\subsubsection{The Invariant and Variant Preference Modeling.} 
For observed data user $u$ and item $i$, we use $\textbf{u}^{inv}$, $\textbf{i}^{inv}$, $\textbf{i}^{var}$, $\textbf{i}^{var}$ to denote the invariant embedding and variant embedding of user $u$ and item $i$. $\textbf{q}_e$ is the corresponding environment embedding. The invariant preference \begin{math}\textbf{m}_{u,i}\end{math} and variant preference \begin{math}\textbf{n}_{u,i}\end{math}  of the user toward item $i$ can be computed as: 
\begin{align}
    \textbf{m}_{u,i} &= \textbf{u}^{inv} \odot \textbf{i}^{inv}, \\
     \textbf{n}_{u,i} &= \textbf{u}^{var} \odot \textbf{i}^{var} \odot \textbf{q}_e,
\end{align}
\label{eq:n}
where $\odot$ is the Hadamard product of two embeddings.

In order to enhance the recommendation abilities of invariant preference, we use the observed feedback label $y$ as the supervision information, and the invariant representations are optimized by:
\begin{equation}
L_{inv} = \frac{1}{|D|}\sum_{(u,i)\in D}l_{rec}(\varphi(\textbf{m}_{u,i}),y_{u,i}),
\label{eq:Linv}
\end{equation}
where $y_{u, i}$ is the user feedback in observed data, $l_{rec}(\cdot)$ is the binary cross entropy loss and $\varphi( \cdot )$ is a predictive function as follow:
\begin{equation}
  \varphi(\textbf{x}) =  Sigmoid(\textbf{x}^T\textbf{1}_L),
  \label{eq:phi}
\end{equation} where $\textbf{1}_L$ is an all-ones vector with the same dimension as \textbf{x}.

\subsubsection{The Adversarial Learning with the Environment Classifier.}
The invariant preference of a user towards an item $\textbf{m}_{u, i}$ tries to confuse the environment classifier $\phi_c$, while the classifier tries to identify the environment label of this pair according to its invariant preference. The ground truth environment label  $e_{u, i}$ can be assigned according to the prediction loss of the observed user feedback (see in Eq.\ref{eq:environment}), since the smaller difference between observed biased feedback and prediction is, the more possibility the pair is from this environment.
The optimization of the environment classifier $\phi_c$ can be defined as:
\begin{equation}
 L_{env} =  \frac{1}{|D|}\sum_{(u,i)\in D}l_{c}(\phi_c(\textbf{m}_{u,i}),e_{u,i}),
\label{eq:e}
\end{equation}
where $\phi_c$ is a single-layer MLP network whose output dimensions are equal to the number of environments, and $l_c$ is the cross entropy loss for multiple classifications.

The environment label $e_{u, i}$ for user-item pair $(u, i)$ is reallocated according to the prediction loss $L_{var}$ of the observed feedback label, formulated as:
\begin{equation}
  e_{u,i} \leftarrow \mathop{\arg \min} \limits_{e \in \mathcal{E}} (L_{var}),
  \label{eq:environment}
\end{equation}
\begin{equation}
    L_{var} = l_{rec}(f(\varphi(\textbf{n}_{u,i}),\varphi(\textbf{m}_{u,i})),y_{u,i}),
\end{equation} 
where $y_{u, i}$ is the biased label in observed data, $\varphi$ is the predictive function (defined in Eq.\ref{eq:phi}), $f(\cdot, \cdot)$ is a product function to fuse the invariant and variant predictions, and $l_{rec}(\cdot)$ is the binary cross entropy loss.

\subsubsection{Overall Objective Functions.}
The goal of invariant preference 
$\textbf{m}_{u, i}= \textbf{u}^{inv} \odot \textbf{i}^{inv} $ is to confuse the environment classifier $\phi_c$, i.e., maximize $L_{env}$,  while the aim of the classifier is to identify the label of the environment, i.e.,  minimize the parameters $\theta_c$ in $L_{env}$. The adversarial process can be formulated as:
\begin{equation}
   \min \limits_{\theta_c} \max \limits_{\mathbf{u}^{inv},\mathbf{i}^{inv}} L_{env}. 
\end{equation}

The overall loss function of disentangled preference modeling can be formulated as:
\begin{equation}
L_{major} = L_{inv} - \alpha \cdot L_{env} + \beta \cdot L_{var},
\end{equation} 
where $\alpha, \beta$ are the hyper-parameters to balance the proportions of the variant and invariant preferences.

\subsection{Distance-aware Unbiased Knowledge Distillation}
Utilizing the Disentangled Preference Modeling module, we systematically decompose user preferences into two distinct categories: invariant and variant preferences.
To achieve unbiased recommendations, we propose a distance-aware knowledge distillation approach.
By conducting the distance-aware knowledge distillation, we distill useful information from both
invariant and variant preference into a lightweight student model, which improves the model efficiency by reducing the number of parameters. Besides, the use of soft predicted labels from the teacher model further enhances the model’s generalization capability.

\subsubsection{Basic Summarization.} 
Invariant preferences represent stable aspects of users' tastes unaffected by external factors, reflecting enduring interests or values. In contrast, variant preferences fluctuate in response to external influences such as time and context. 
By adaptively integrating these two preferences, recommender systems can better understand and anticipate user needs, ultimately facilitating unbiased recommendations.
Our method is grounded in a fundamental principle: when there is a significant difference in the invariant and variant preferences, it indicates that the user's current needs diverge markedly from their long-term preferences. By incorporating variant preferences, the system can effectively capture this contextual information, allowing it to deliver recommendations that are not only more relevant but also unbiased. 
In practice, this means that the greater the discrepancy between the variant preference and invariant preference, the more attention should be paid to the information in the variant preference. Conversely, when the discrepancy is minimal, such imputation information may be less required.
Ultimately, we derive users' invariant and variant-dominated feedback, measuring their differences through distance calculations. This difference informs the adaptive fusion of preferences, which serves as soft labels to guide the training of the student model. 
\subsubsection{Teacher Model.} 
Given a user-item pair $(u,i)$, we can calculate the invariant-dominated feedback \begin{math}\hat{y}_{u, i}^{inv}\end{math} and the variant-dominated feedback \begin{math}\hat{y}_{u, i}^{var}\end{math} respectively:
\begin{align}
    \hat{y}_{u,i}^{inv} &= \varphi(\textbf{m}_{u,i}), \\
    \hat{y}_{u,i}^{var} &= \varphi(\textbf{n}_{u,i}).
\end{align}
where $\varphi$ is the predictive function (defined in Eq.\ref{eq:phi}).

Then the distance between the prediction labels of invariant and variant preferences is calculated as an indicator to fuse the prediction results. The fusion process in the teacher model can be formulated as follows:
\begin{align}
    d_{u,i} &=  |\hat{y}_{u,i}^{inv} - \hat{y}_{u,i}^{var}|, \\
    \label{eq:gamma}
    w_{u,i}^{inv} &= (1 - d_{u,i} )^\gamma, \\
    w_{u,i}^{var} &= 1 - w_{u,i}^{inv}, \\
    \hat{y}_{u,i}^{*} &= w_{u,i}^{inv}*\hat{y}_{u,i}^{inv} + w_{u,i}^{var}*\hat{y}_{u,i}^{var},
    \label{eq:y_star}
\end{align}
where $\gamma$ is a hyper-parameter and acts as the temperature coefficient in the fusion process. 
With the increasing of $\gamma$, more information on variant preferences be used to predict the final unbiased user feedback. 

\subsubsection{Student Model.}
In order to enhance model performance and facilitate downstream recommendation tasks, we introduce a distance-aware knowledge distillation strategy to distill the unbiased information from the teacher model to a lightweight student model. 
Specifically, a simple matrix factorization-based student model is
designed to capture the collaborative signals. The prediction label $\hat{y}^{stu}_{u,i}$ of the student model is defined as:
\begin{equation}
\hat{y}^{stu}_{u,i}=\textbf{s}_{u,i} \odot \textbf{t}_{u,i},
\end{equation}
where $\textbf{s}_{u, i}$ and $\textbf{t}_{u, i}$ are the learning representations of the user and item in the student model.

For each user-item pair (\emph{u, i}) in observed data, we treat the predicted unbiased user feedback  $\hat{y}_{u,v}^{*}$ in the teacher model (see in Eq.\ref{eq:y_star}) as the ground truth label.
The loss function in knowledge distillation process is defined as:
\begin{equation}
    L_{KD} = \frac{1}{|D|}\sum_{(u,i)\in D}l_{rec}(\hat{y}^{stu}_{u,i}, \hat{y}_{u,i}^{*}),
\end{equation}
where $l_{rec}(\cdot)$ is the binary cross-entropy loss.

In summary, we first disentangle the variant and invariant preference of users, then propose a student model to make the distillation of fused information, which not only lightweight the model parameters but also increases the generalization ability of the debiasing student model.
Without the requirement of unbiased data, our debiasing model is trained in an unsupervised way, which expands the applicability of KD-Debias.

\section{Experiments}

We conduct experiments on three public datasets and compare our model with representative debiasing models to show the effectiveness of KD-Debias.
\subsection{Experimental Settings}

\begin{table}
\caption{Statistics of datasets.}
\label{Statistics of the datasets.}
\begin{tabular}{ccccccc}
\hline
Dataset   & \#Users  & \#Item    & Biased Data   & Unbiased Data \\ \hline
Yahoo!R3  & 15.4k  & 1.0k      & 250k          & 54k   \\
Coat      & 290    & 300       & 6.9k          & 4.2k  \\
Mind      & 100k   &  161k     & 24M           & -     \\ \hline
\end{tabular}
\end{table}
\subsubsection{Datasets.}
To validate the effectiveness of KD-Debias, we utilize three real-world datasets: Yahoo!R3, Coat, and Mind datasets. The statistics of three datasets are shown in Table~\ref{Statistics of the datasets.}.
\begin{itemize}
    \item \textbf{Yahoo!R3}\footnote{https://webscope.sandbox.yahoo.com/catalog.php?datatype=r\&did=3}: The Yahoo! R3 dataset is a collection of user ratings for songs, collected within the context of music recommendation. The rating scale range from 1 to 5. It consists of two subsets: a set of biased data collected from norm recommendation scenarios, and a set of unbiased data collected by a random logging strategy where items are posted randomly. 
    \item\textbf{Coat}\footnote{https://www.cs.cornell.edu/~schnabts/mnar/}: The dataset is sourced from a coat recommendation platform, capturing user ratings for items on a scale of 1 to 5. It consists of two subsets:
    a set of biased data collected from norm recommendation scenarios, and a set of unbiased data collected by a random logging strategy where items are posted randomly. 
    \item\textbf{Mind}\footnote{https://paperswithcode.com/dataset/mind}: is a widely used news dataset and it contains user clicks for news and which news is exposed to users. It is widely used to study the debiasing problem of exposure bias.
\end{itemize}

Following the experimental settings in~\citep{ding2022interpolative,wang2022invariant}, the biased data is used in the training process. For the methods under the supervision of unbiased data, we randomly sample 5\% of unbiased data as the supervised labels in the training set (this part is not used in unsupervised methods), 5\% of unbiased data as the validation set for choosing parameters, the remaining 90\% of the unbiased data is used for testing. The rating scores (1-5) larger than 3 are treated as positive feedback, while the other rating scores are treated as negative feedback.

\subsubsection{Evaluation metrics.}
We adopt the NDCG$@k$ and Recall$@k$ to evaluate the performance of our model. 
\begin{itemize}
\item \textbf{NDCG$@k$}: measures the quality of recommendations by combining relevance and position information.
\begin{align}
    DCG@k&= \sum_{(u, i)} \frac{\mathbb{I}(\hat{z}_{u, i} \leq k)}{log(\hat{z}_{u, i} + 1)},\\
    NDCG@k &= \frac{1}{|\mathcal U|}\sum_{u \in \mathcal U} \frac{DCG@k}{IDCG@k},
\end{align} 
where $\hat{z}_{u, i}$ is the rank position of item \emph{i} in the recommended list for user \emph{u}, IDCG$@k$ is the ideal DCG$@k$. $\mathbb{I} (\hat{z}_{u, i} \leq k)$ returns 1 if $\hat{z}_{u, i} \leq k$, and 0 otherwise.

\item \textbf{Recall$@k$}: measures the number of items appear in the top-K clicked list of users.
\begin{align}
    Recall_u@k&= \frac{\sum_{(u,i)}\mathbb{I}(\hat{z}_{u,i} \leq k)}{|I_u|}, \\
    Recall@k &= \frac{1}{|\mathcal U|}\sum_{u \in \mathcal U} Recall_u@k,
\end{align} where $I_u$ is the item set of all interactions of user u.
\end{itemize}

\subsubsection{Baselines.}
We compare our proposed framework KD-Debias with the mainstream debiasing methods in recommender systems.
\begin{itemize}
\item \textbf{MF}~\citep{koren2009matrix}: Matrix Factorization is a widely used model in the recommendation, which is a biased model without debiasing capability.
\item \textbf{MF-IPS}~\citep{schnabel2016recommendations}: is a recommendation model that incorporates inverse propensity scores into the MF framework to address bias issues.
\item \textbf{CauseE}~\citep{bonner2018causal}:  is a popular embedding method that is able to predict unbiased user feedback by using biased data and unbiased data.
\item \textbf{AutoDebias}~\citep{chen2021autodebias}: is a debiasing model trained with normal biased and unbiased data by utilizing meta-learning.
\item \textbf{InterD}~\citep{ding2022interpolative}: is a SOTA debiasing knowledge distillation method in recommender systems, interD takes MF and AutoDebias as biased-teacher and unbiased teacher to distill a student model.
\item \textbf{InvPref}~\citep{wang2022invariant}:  is a SOTA unsupervised debiasing method in recommender systems. InvPref treats environments as the agent of confounders and applies invariant learning to learn the user’s unbiased preference.
\end{itemize}

To further verify the debiasing ability of our model on a specific bias, we select one of the most common biases in recommender systems, i.e., the exposure bias. In addition to comparing with the general debiasing methods, we also compare our method with the representative methods that specially designed for exposure bias, including:
\begin{itemize}
\item \textbf{WMF}~\citep{hu2008collaborative}: is the most representative debiasing model for exposure bias.  WMF reduces exposure bias by imputing and assigning lower weights to unobserved samples. 
\item \textbf{EXMF}~\citep{liang2016modeling}: it models the probability of exposure and converted it as confidence degree weight to reduce exposure bias.
\end{itemize}

The above methods cover different kinds of debiasing approaches in
recommender systems: 
MF-IPS and AutoDebias leverage IPS and meta-learning techniques for debiasing. CauseE and InterD are knowledge distillation debiasing approaches with lightweight unbiased models. All of them need the supervision of unbiased data.
The most similar work to our proposed framework is InvPref, which conducts unbiased recommendations without relying on unbiased data. Different from InvPref, our KD-Debias framework is a general debiasing framework, which automatically incorporates both invariant and variant information via a distance-aware knowledge distillation process. Table~\ref{tb-baseline} summarizes the properties of different methods.

\begin{table}[h]
\centering
\caption{Properties of debiasing methods. Debiasing: a debiasing model?; Unsupervised: training with unbiased data? General: a general debiasing model?  KD: knowledge distillation model?}
\label{tb-baseline}
\begin{tabular}{ c c c c c} \hline
~ & Debiasing & Unsupervised & General & KD \\ \hline 
MF & $\times$  & $\surd$ & $\times$ &$\times$   \\
MF-IPS &  $\surd$ &$\times$ & $\surd$  &$\times$  \\
AutoDebias &  $\surd$ &$\times$ & $\surd$  &$\times$  \\
CauseE &  $\surd$ &$\times$ & $\surd$  &$\surd$  \\
InterD &  $\surd$ &$\times$ & $\surd$  &$\surd$  \\
InvPref &  $\surd$ &$\surd$ & $\surd$  &$\times$  \\
WMF &  $\surd$ &$\surd$ & $\times$  &$\times$  \\
EXMF &  $\surd$ &$\surd$ & $\times$  &$\times$  \\
\textbf{KD-Debias} &  $\surd$ &$\surd$ & $\surd$  &$\surd$  \\\hline
\end{tabular}
\end{table}

\subsubsection{Parameter Settings. }
For each baseline method, a grid search is applied to find the optimal settings. These include embedding dimensions from $\{10,20,30
,40,50\}$, and the learning rate from $\{0.1, 0.01, 0.001, 0.0001, 0.00001\}$. We report the result of each method with its optimal hyperparameter settings on the validation data. 
In our KD-Debias model, we set the dimension of invariant and variant embedding to 40, the number of environments is 2, the learning rate in disentangled preference modeling is $0.003$, the learning rate in distance-aware distillation is $0.005$, the $\alpha$, $\beta$, $\gamma$ are set to 1.9, 9.9, 0.17.
The code will be publicly available after the review process.

\subsection{Main Results}
We compare the performances of methods in
the general debiasing and specific debiasing (i.e., exposure bias) settings. As shown in Table~\ref{Performace Comparison on yahoo}, Table~\ref{Performace Comparison on coat} and Table~\ref{Exposure Bias Comparison}, we can observe that: 

(1) MF performs the worst in general debiasing test due to it is not a debiased model. Compared to MF, MF-IPS addressing specific biases without uniform data can bring benefits.

(2) CauseE, AutoDebias, and InterD steadily outperform MF and MF-IPS. Under the supervision of unbiased data, these methods gain improvements in resisting affect from biases.

(3) The unsupervised model InvPerf outperforms other baseline methods in Yahoo!R3 dataset, suggesting that the invariant preference can reveal the unbiased preference of users to some extent. However, due to the information loss problem, InvPerf loses the advantages of the supervised method InterD in the Coat dataset.

(4) Our proposed method KD-Debias achieves the best performance on all datasets, showing the effectiveness of our model by incorporating the biased information in the distance-aware knowledge distillation framework.  
(5) For the model efficiency of the training methods with observed biased data only, our surrogate student model in KD-Debias is a lightweight model, which achieves the best performance with less than $50\%$ learning parameters compared to the SOTA baseline method InvPerf.

(6) For the specific exposure bias, we report the NDCG and Recall at top $K=10,20,40$ in the rank lists. As shown in Table~\ref{Exposure Bias Comparison}, our model achieves the best performance on both NDCG and Recall in terms of addressing exposure bias.
This observation indicates that with the imputation of biased information, our knowledge distillation method learns more useful information for unbiased recommendations.

\begin{table*}[]
\centering
\caption{Performace comparison on Yahoo!R3 dataset. The improvements achieved by KD-Debias are statistically significant (i.e., two-sided t-test with p-value < 0.05) over the best baseline (marked by underline). 
}
\label{Performace Comparison on yahoo}
{
\begin{tabular}{|c|ccccc|}
\hline
\multirow{2}{*}{Method} & \multicolumn{5}{c|}{Yahoo!R3}           \\ \cline{2-6}
                        & \multicolumn{1}{c|}{NDCG$@$5}  & \multicolumn{1}{c}{Imp.$\uparrow$}     & \multicolumn{1}{|c|}{Recall@5} & \multicolumn{1}{c}{Imp.$\uparrow$}   & \multicolumn{1}{|c|}{\# Parameters}        \\ \hline
MF         & \multicolumn{1}{c|}{0.6007}  & \multicolumn{1}{c|}{-}     & \multicolumn{1}{c|}{0.7639}   & \multicolumn{1}{c|}{-}   & 492k\\
MF-IPS     & \multicolumn{1}{c|}{0.6113}  & \multicolumn{1}{c|}{+1.76\%}      & \multicolumn{1}{c|}{0.7754}   & \multicolumn{1}{c|}{+1.51\%}  & 492k\\
CauseE     & \multicolumn{1}{c|}{0.6108}  & \multicolumn{1}{c|}{+1.68\%}      & \multicolumn{1}{c|}{0.7694}   & \multicolumn{1}{c|}{+0.72\%}  & 492k\\
AutoDebias & \multicolumn{1}{c|}{0.6134}   & \multicolumn{1}{c|}{+2.11\%}     & \multicolumn{1}{c|}{0.7762}   & \multicolumn{1}{c|}{+1.61\%}  & 180k\\
InterD    & \multicolumn{1}{c|}{0.6263}    & \multicolumn{1}{c|}{+4.26\%}       & \multicolumn{1}{c|}{0.7853}  & \multicolumn{1}{c|}{+2.80\%}   & 180k\\
InvPref    & \multicolumn{1}{c|}{{\ul 0.6578}} & \multicolumn{1}{c|}{+9.51\%}  & \multicolumn{1}{c|}{{\ul 0.8065}} & \multicolumn{1}{c|}{+5.58\%} & {\ul 
 1.31M}\\
KD-Debias & \multicolumn{1}{c|}{\textbf{0.6739$^{*}$}} & \multicolumn{1}{c|}{\textbf{+12.19 \%}}  & \multicolumn{1}{c|}{\textbf{0.8217$^{*}$}} & \multicolumn{1}{c|}{\textbf{+7.57\%}} & \textbf{656k}\\ \hline
\end{tabular}
}
\end{table*}

\begin{table*}[]
\centering
\caption{Performace comparison on Coat dataset. The improvements achieved by KD-Debias are statistically significant (i.e., two-sided t-test with p-value < 0.05) over the best baseline (marked by underline). 
}
\label{Performace Comparison on coat}
{
\begin{tabular}{|c|ccccc|}
\hline
\multirow{2}{*}{Method} & \multicolumn{5}{c|}{Coat}           \\ \cline{2-6}
                        & \multicolumn{1}{c|}{NDCG@5}  & \multicolumn{1}{c}{Imp. $\uparrow$} & \multicolumn{1}{|c|}{Recall@5} & \multicolumn{1}{c}{Imp. $\uparrow$} & \multicolumn{1}{|c|}{\# Parameters}         \\ \hline
MF        & \multicolumn{1}{c|}{0.4128} & \multicolumn{1}{c|}{-}  & \multicolumn{1}{c|}{0.4716}  & \multicolumn{1}{c|}{-} & 17.7k          \\
MF-IPS     & \multicolumn{1}{c|}{0.4131}  & \multicolumn{1}{c|}{+0.07\%} & \multicolumn{1}{c|}{0.4775}  & \multicolumn{1}{c|}{+1.25\%} & 17.7k          \\
CauseE     & \multicolumn{1}{c|}{0.4166} & \multicolumn{1}{c|}{+0.92\%}  & \multicolumn{1}{c|}{0.4734} & \multicolumn{1}{c|}{+0.38\%}  & 17.7k          \\
AutoDebias & \multicolumn{1}{c|}{0.4380} & \multicolumn{1}{c|}{+6.10\%}  & \multicolumn{1}{c|}{0.4899} & \multicolumn{1}{c|}{+3.88\%}  & 6.5k \\
InterD    & \multicolumn{1}{c|}{{\ul 0.4475}}  & \multicolumn{1}{c|}{+8.41\%}   & \multicolumn{1}{c|}{{\ul 0.5121}} & \multicolumn{1}{c|}{+8.59\%}   & 6.5k \\
InvPref    & \multicolumn{1}{c|}{0.4241} & \multicolumn{1}{c|}{+2.74\%} & \multicolumn{1}{c|}{0.4721} & \multicolumn{1}{c|}{+0.11\%}  & {\ul 47.3k}        \\
KD-Debias  & \multicolumn{1}{c|}{\textbf{0.4687$^{*}$}} & \multicolumn{1}{c|}{\textbf{+13.54\%}}  & \multicolumn{1}{c|}{\textbf{0.5180$^{*}$}} & \multicolumn{1}{c|}{\textbf{+9.83\%}} & \textbf{23.6k}          \\ \hline
\end{tabular}
}
\end{table*}

\begin{table}[t]
\centering
\caption{Performace comparisons in dealing with exposure bias on MIND dataset.}
\label{Exposure Bias Comparison}
\begin{tabular}{|c|cccccc|}
\hline
\multirow{2}{*}{Method} 
                        & \multicolumn{3}{c|}{NDCG}                                                                                       & \multicolumn{3}{c|}{Recall}                                                                \\ \cline{2-7} 
                        & \multicolumn{1}{c|}{Top10}          & \multicolumn{1}{c|}{Top20}          & \multicolumn{1}{c|}{Top40}          & \multicolumn{1}{c|}{Top10}          & \multicolumn{1}{c|}{Top20}          & Top40          \\ \hline
MF                      & \multicolumn{1}{c|}{0.222}          & \multicolumn{1}{c|}{0.258}          & \multicolumn{1}{c|}{0.283}          & \multicolumn{1}{c|}{0.362}          & \multicolumn{1}{c|}{0.499}          & 0.591          \\
WMF                     & \multicolumn{1}{c|}{0.234}          & \multicolumn{1}{c|}{0.267}          & \multicolumn{1}{c|}{0.301} & \multicolumn{1}{c|}{0.379}          & \multicolumn{1}{c|}{0.513}          & 0.607          \\
EXMF                    & \multicolumn{1}{c|}{0.228}          & \multicolumn{1}{c|}{0.264}          & \multicolumn{1}{c|}{0.289}          & \multicolumn{1}{c|}{0.362}          & \multicolumn{1}{c|}{0.499}          & 0.588          \\
InvPref                 & \multicolumn{1}{c|}{0.234}          & \multicolumn{1}{c|}{0.270}          & \multicolumn{1}{c|}{0.298}          & \multicolumn{1}{c|}{0.386} & \multicolumn{1}{c|}{0.532}          & 0.624          \\
KD-Debias               & \multicolumn{1}{c|}{\textbf{0.238}} & \multicolumn{1}{c|}{\textbf{0.278}} & \multicolumn{1}{c|}{\textbf{0.301}} & \multicolumn{1}{c|}{\textbf{0.386}} & \multicolumn{1}{c|}{\textbf{0.545}} & \textbf{0.661} \\ \hline
\end{tabular}
\label{Exposure Bias Comparison}
\end{table}

\subsection{Experimental Analysis}
We conduct in-depth analyses of our model to quantitatively show the effectiveness and stability of KD-Debias.

\subsubsection{Ablation Study.}
To demonstrate the effectiveness of each component (including the variant information, the distance-aware fusion strategy and the knowledge distillation process) in our proposed KD-Debias, we conduct the ablation study on Yahoo!R3 and Coat datasets. The degeneration versions of KD-Debias are as follows: 

\begin{itemize}
\item{Ours-Variant}: it only utilizes invariant user preference to guide the learning of student model. 
\item{Ours-Distance}: it replaces the distance-aware combination with a simple combination strategy ( i.e., information are fused with the equal weight 0.5) of the invariant and variant preference in the knowledge distillation process. 
\item {Ours-KD}: the invariant and variant preference are fused with distance-aware mode for  unbiased recommendations without knowledge distillation process. The teacher model is used to make prediction for unbiased recommendations. 
\end{itemize}

As shown in Table~\ref{Ablation Study}, we can observed that:
(1) The performance of Ours-Variant is worse than KD-Debias on all metrics, which verifies the effectiveness of incorporating biased information. (2) Replacing the fusion function (Ours-Distance) will result in the degeneration of model performance, showing the usefulness of our proposed distance-aware module. This emphasizing the critical role of effective information fusion in enhancing overall model performance.
(3) By removing the knowledge distillation module (Ours-KD), although the variant and invariant preferences are intuitively fused with weighted coefficients, the model performance decreases a lot, showing the necessity to design a knowledge distillation framework with the surrogate student model to fuse information. This emphasizes the importance of knowledge distillation in effectively blending diverse information for better performance.

\begin{table}
\centering
\caption{The performance of the degeneration versions of KD-Debias.}
\label{Ablation Study}
\begin{tabular}{|c|cc|cc|}
\hline
Dataset & \multicolumn{2}{c|}{Yahoo!R3}             & \multicolumn{2}{c|}{Coat}                \\ \hline
Metric  & \multicolumn{1}{c|}{NDCG@5} & Recall@5 & \multicolumn{1}{c|}{NDCG@5} & Recall@5 \\ \hline
Ours  & \multicolumn{1}{c|}{\textbf{0.6739}} & {\textbf{0.8217}} & \multicolumn{1}{c|}{\textbf{0.4687}}   & {\textbf{0.5180}}  \\ 
Ours-Variant   & \multicolumn{1}{c|}{0.6693} & {0.8125} & \multicolumn{1}{c|}{0.4278}   & {0.4695} \\
Ours-Distance   & \multicolumn{1}{c|}{0.6596} & {0.8089} & \multicolumn{1}{c|}{0.4643}   & {0.5142} \\
Ours-KD   & \multicolumn{1}{c|}{0.6326} & {0.8046} & \multicolumn{1}{c|}{0.4291}   & {0.4723} \\ \hline
\end{tabular}
\end{table}

\begin{figure}
    \centering
    \subfloat[NDCG@5]{\includegraphics[width=0.4\linewidth]{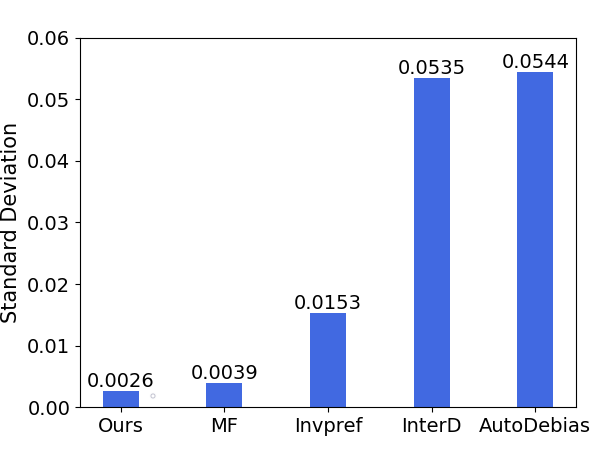}}
    \subfloat[Recall@5]{\includegraphics[width=0.4\linewidth]{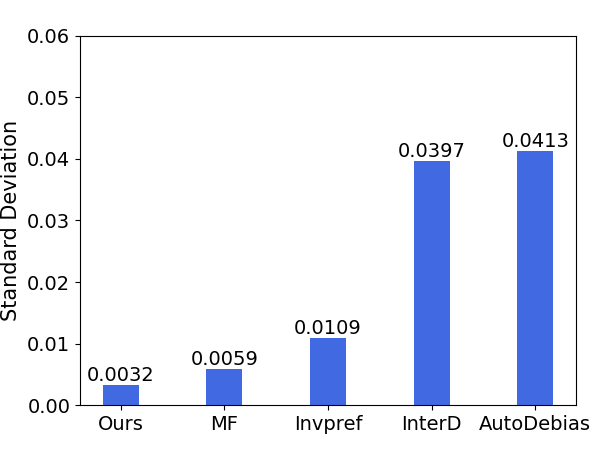}}
    \caption{Stability comparison on Yahoo!R3 dataset. The standard deviation is computed based on 20 times repeated experiments with different random seeds.}
    \label{fg:Stablility}
\end{figure}

\subsubsection{The Stability Analysis.}
We conducted 20 repeated experiments with different random seeds in the same settings to evaluate the stability of models. We compute the standard deviation of the performance on each model on Yahoo!R3 dataset, and choose the base model (i.e., MF~\citep{koren2009matrix} ) in these methods as the baseline to compare with.
As shown in Fig.~\ref{fg:Stablility}, there is a certain gap between InvPref and the base model MF, while the standard deviations of the model performance in InterD and AutoDebias are 13 times larger than MF, showing the instability of these methods.
The stability of our KD-Debias is comparable to the base model MF. It may benefit from the distance-aware knowledge distillation module, which combines invariant preference and variant preference by considering the inconsistency distribution between them and meanwhile increases the generalization ability via learning the soft predicted labels from the teacher model.

\begin{figure}
    \centering
    \subfloat[Yahoo!R3 Dataset]{\includegraphics[width=0.4\linewidth]{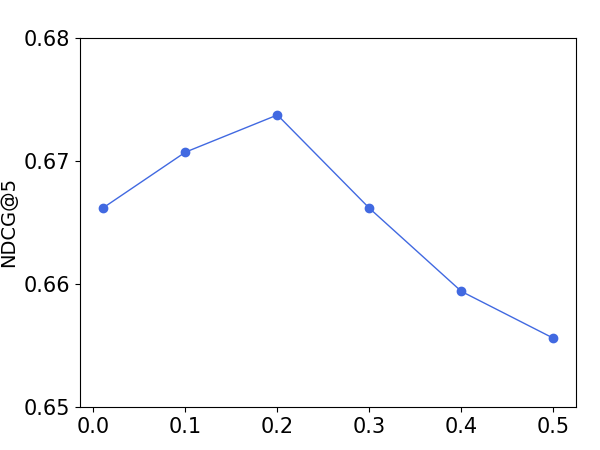}}
    \subfloat[Coat Dataset]{\includegraphics[width=0.4\linewidth]{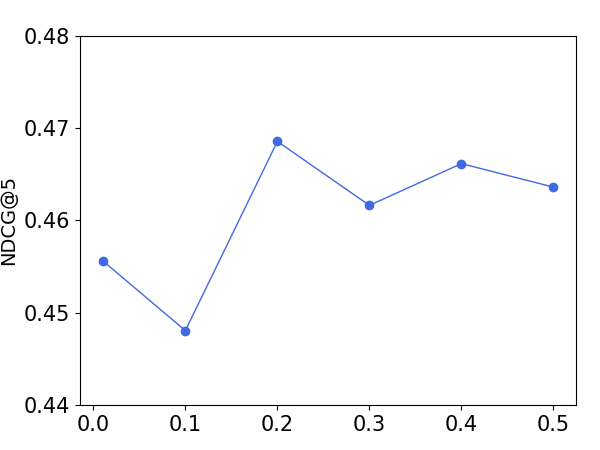}}
    \caption{The performances comparison of NDCG@5 with different hyper-parameter $\gamma$.}
    \label{fg:gama}
\end{figure}

\begin{figure}
    \centering
    \subfloat[Yahoo!R3 Dataset]{\includegraphics[width=0.4\linewidth]{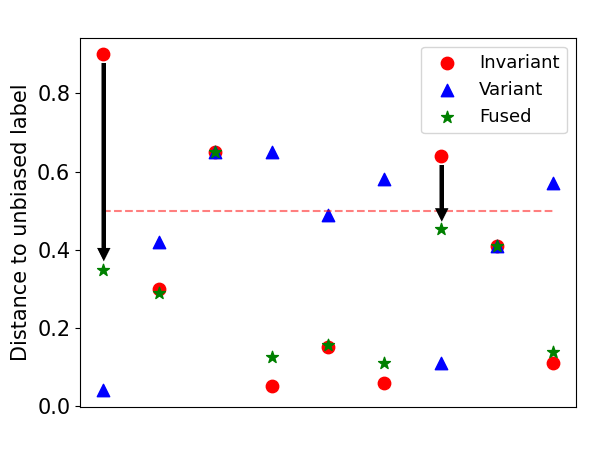}}
    \subfloat[Coat Dataset]{\includegraphics[width=0.4\linewidth]{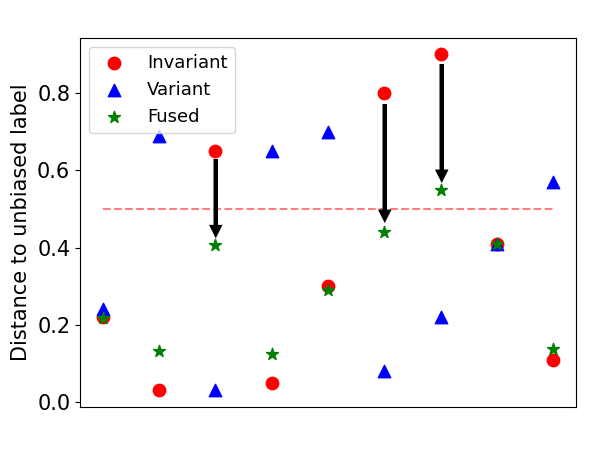}}
    \caption{The distance between the prediction results of invariant preferences (marked by the red circles), variant preferences (marked by the blue triangles), and the fused preference (marked by the green stars). The dots under the red dashed line (y = 0.5) can be regarded as correct predictions.}
    \label{fig:fuse}
\end{figure}

\subsubsection{Hyperparameters Analysis.}
We conduct analysis of the hyperparameters (e.g. the temperature coefficient $\gamma$) that may influence the model performance.
As mentioned in Eq.13, $\gamma$ is a hyper-parameter acting as the temperature co-efficient in knowledge distillation to smooth the fusion of unbiased information and biased information. As shown in Fig. \ref{fg:gama}, as the increasing of $\gamma$, $NDCG@5$ has a clear trend of increasing and then decreasing. The increasing part indicated that incorporating biased information enhances the performance of unbiased recommendations, while incorporating too much variant information may confound invariant preference and substantially decrease the performance of unbiased recommendations.

\subsubsection{Case Study.}
We have demonstrated the effectiveness of incorporating variant information in knowledge distillation. We intuitively display its effect on the model performance in Fig.~\ref{fig:fuse}. The value on the vertical axis represents the distance to the true unbiased label.  The smaller value is, the more precise of prediction results. 
We can see that variant information (marked by blue triangles) plays a complementary role to invariant information (marked by the red circle) in debiasing learning, and their fusion makes the prediction results (marked by green stars) closer to the true unbiased label. 
It shows the the effectiveness of variant preferences in our model in enhancing the debiasing performance in recommender systems.

\section{Conclusion}
In this paper, we investigate the problem of general debiasing with only the biased observation data.
We point out the instability and low accuracy of invariant learning in unbiased recommendations, and propose a novel distance-aware knowledge distillation framework (KD-Debias) to make unbiased predictions with both invariant and variant information. 
Extensive experiments show that with the imputation from variant information, our knowledge distillation method achieves the best model performance in unbiased recommendation task. 
In our framework, a simple MF method is used as the student model to improve the model efficiency and generalization ability in unbiased recommendations.
In the future, we will expand the applications of our framework with more complex student models in different recommendation scenarios.

\bibliographystyle{model1-num-names.bst}
\bibliography{mybib}


\end{document}